\documentstyle[12pt]{article}
\newlength{\extraspace}
\newlength{\extraspaces}
\catcode`\@=11
\def\numberbysection{\@addtoreset{equation}{section}
\def\theequation{\arabic{section}.\arabic{equation}}}
\newcommand{\be}{\begin{equation}
\addtolength{\abovedisplayskip}{\extraspaces}
\addtolength{\belowdisplayskip}{\extraspaces}
\addtolength{\abovedisplayshortskip}{\extraspace}
\addtolength{\belowdisplayshortskip}{\extraspace}}
\newcommand{\ee}{\end{equation}}
\newcommand{\ba}{\begin{eqnarray}
\addtolength{\abovedisplayskip}{\extraspaces}
\addtolength{\belowdisplayskip}{\extraspaces}
\addtolength{\abovedisplayshortskip}{\extraspace}
\addtolength{\belowdisplayshortskip}{\extraspace}}
\newcommand{\ea}{\end{eqnarray}}
\newcommand{\nonu}{\nonumber \\[.5mm]}
\newcommand{\A}{&\!\!\!}

\newcommand{\e}{\, {\rm e}}

\newcommand{\X}{{\hat X}}

\newcommand{\VEV}[1]{\left\langle {#1} \right\rangle}

\newcommand{\figone}{
\begin{figure}[tb]
\setlength{\unitlength}{0.0125in}
\begin{picture}(280,85)(25,745)
\thicklines
\put(110,800){\circle{40}}
\put( 90,800){\line(-1, 0){ 20}}
\put( 70,805){\makebox(0,0)[lb]
    {\raisebox{0pt}[0pt][0pt]{\twlrm $R$}}}
\put(105,755){\makebox(0,0)[lb]
    {\raisebox{0pt}[0pt][0pt]{\twlrm (a)}}}
\put(260,800){\circle{40}}
\put(240,800){\line(-1, 0){ 20}}
\put(220,805){\makebox(0,0)[lb]
    {\raisebox{0pt}[0pt][0pt]{\twlrm $\Omega^2$}}}
\put(255,755){\makebox(0,0)[lb]
    {\raisebox{0pt}[0pt][0pt]{\twlrm (b)}}}
\put(410,800){\circle{40}}
\put(390,800){\line(-1, 0){ 20}}
\put(430,800){\line( 1, 0){ 20}}
\put(370,805){\makebox(0,0)[lb]
    {\raisebox{0pt}[0pt][0pt]{\twlrm $\Omega$}}}
\put(440,805){\makebox(0,0)[lb]
    {\raisebox{0pt}[0pt][0pt]{\twlrm $\Omega$}}}
\put(405,755){\makebox(0,0)[lb]
    {\raisebox{0pt}[0pt][0pt]{\twlrm (c)}}}
\end{picture}
\caption{One-loop divergent diagrams without quantum gravity.}
\label{figureone}
\vspace{5mm}
\end{figure}}
\newcommand{\figtwo}{
\begin{figure}[tb]
\setlength{\unitlength}{0.0125in}
\begin{picture}(280,85)(-40,745)
\thicklines
\put(100,800){\circle{40}}
\put( 80,800){\line(-1, 0){ 20}}
\put( 60,805){\makebox(0,0)[lb]
    {\raisebox{0pt}[0pt][0pt]{\twlrm $G$}}}
\put(123,795){\makebox(0,0)[lb]
    {\raisebox{1pt}[0pt][0pt]{\twlrm $h$}}}
\put( 95,755){\makebox(0,0)[lb]
    {\raisebox{0pt}[0pt][0pt]{\twlrm (a)}}}
\put(260,800){\circle{40}}
\put(240,800){\line(-1, 0){ 20}}
\put(280,800){\line( 1, 0){ 20}}
\put(220,805){\makebox(0,0)[lb]
    {\raisebox{0pt}[0pt][0pt]{\twlrm $E$}}}
\put(290,805){\makebox(0,0)[lb]
    {\raisebox{0pt}[0pt][0pt]{\twlrm $E$}}}
\put(257,823){\makebox(0,0)[lb]
    {\raisebox{1pt}[0pt][0pt]{\twlrm $h$}}}
\put(255,755){\makebox(0,0)[lb]
    {\raisebox{0pt}[0pt][0pt]{\twlrm (b)}}}
\end{picture}
\caption{$O(\kappa_0^2)$ divergent diagrams.
Internal lines with $h$ are graviton propagators.
Other internal lines are matter propagators.}
\label{figuretwo}
\vspace{5mm}
\end{figure}}
\newcommand{\figthree}{
\begin{figure}[tb]
\setlength{\unitlength}{0.0125in}
\begin{picture}(460,295)(30,535)
\thicklines
\put(110,800){\circle{40}}
\put( 90,800){\line(-1, 0){ 20}}
\put( 90,800){\line(1,0){40}}
\put( 70,805){\makebox(0,0)[lb]
    {\raisebox{0pt}[0pt][0pt]{\twlrm $R$}}}
\put(107,803){\makebox(0,0)[lb]
    {\raisebox{0pt}[0pt][0pt]{\twlrm $h$}}}
\put(105,755){\makebox(0,0)[lb]
    {\raisebox{0pt}[0pt][0pt]{\twlrm (a)}}}
\put(260,800){\circle{40}}
\put(240,800){\line(-1, 0){ 20}}
\put(260,780){\line(0,1){40}}
\put(220,805){\makebox(0,0)[lb]
    {\raisebox{0pt}[0pt][0pt]{\twlrm $R$}}}
\put(263,795){\makebox(0,0)[lb]
    {\raisebox{0pt}[0pt][0pt]{\twlrm $h$}}}
\put(255,755){\makebox(0,0)[lb]
    {\raisebox{0pt}[0pt][0pt]{\twlrm (b)}}}
\put(390,800){\circle{40}}
\put(430,800){\circle{40}}
\put(370,800){\line(-1, 0){ 20}}
\put(350,805){\makebox(0,0)[lb]
    {\raisebox{0pt}[0pt][0pt]{\twlrm $R$}}}
\put(453,795){\makebox(0,0)[lb]
    {\raisebox{0pt}[0pt][0pt]{\twlrm $h$}}}
\put(405,755){\makebox(0,0)[lb]
    {\raisebox{0pt}[0pt][0pt]{\twlrm (c)}}}
\put( 90,695){\circle{40}}
\put(130,695){\circle{40}}
\put(110,695){\line(0, 1){ 40}}
\put(113,725){\makebox(0,0)[lb]
    {\raisebox{0pt}[0pt][0pt]{\twlrm $R$}}}
\put(153,690){\makebox(0,0)[lb]
    {\raisebox{0pt}[0pt][0pt]{\twlrm $h$}}}
\put(105,650){\makebox(0,0)[lb]
    {\raisebox{0pt}[0pt][0pt]{\twlrm (d)}}}
\put(240,695){\circle{40}}
\put(280,695){\circle{40}}
\put(260,695){\line(0, 1){ 40}}
\put(300,695){\line(1, 0){ 20}}
\put(263,725){\makebox(0,0)[lb]
    {\raisebox{0pt}[0pt][0pt]{\twlrm $R$}}}
\put(310,700){\makebox(0,0)[lb]
    {\raisebox{0pt}[0pt][0pt]{\twlrm $E$}}}
\put(296,710){\makebox(0,0)[lb]
    {\raisebox{0pt}[0pt][0pt]{\twlrm $h$}}}
\put(255,650){\makebox(0,0)[lb]
    {\raisebox{0pt}[0pt][0pt]{\twlrm (e)}}}
\put(390,695){\circle{40}}
\put(430,695){\circle{40}}
\put(410,695){\line(0, 1){ 40}}
\put(444,709){\line(1, 1){ 15}}
\put(444,681){\line(1, -1){ 15}}
\put(413,725){\makebox(0,0)[lb]
    {\raisebox{0pt}[0pt][0pt]{\twlrm $R$}}}
\put(443,725){\makebox(0,0)[lb]
    {\raisebox{0pt}[0pt][0pt]{\twlrm $E$}}}
\put(443,655){\makebox(0,0)[lb]
    {\raisebox{0pt}[0pt][0pt]{\twlrm $E$}}}
\put(453,690){\makebox(0,0)[lb]
    {\raisebox{0pt}[0pt][0pt]{\twlrm $h$}}}
\put(405,650){\makebox(0,0)[lb]
    {\raisebox{0pt}[0pt][0pt]{\twlrm (f)}}}
\put(110,590){\circle{40}}
\put( 90,590){\line(-1, 0){ 20}}
\put( 90,590){\line(1,0){40}}
\put(110,610){\line(0,1){20}}
\put( 70,595){\makebox(0,0)[lb]
    {\raisebox{0pt}[0pt][0pt]{\twlrm $R$}}}
\put(113,620){\makebox(0,0)[lb]
    {\raisebox{0pt}[0pt][0pt]{\twlrm $E$}}}
\put(126,605){\makebox(0,0)[lb]
    {\raisebox{0pt}[0pt][0pt]{\twlrm $h$}}}
\put(105,545){\makebox(0,0)[lb]
    {\raisebox{0pt}[0pt][0pt]{\twlrm (g)}}}
\put(210,590){\circle{40}}
\put(190,590){\circle*{8}}
\put(190,590){\line(-1, 0){ 22}}
\put(168,595){\makebox(0,0)[lb]
    {\raisebox{0pt}[0pt][0pt]{\twlrm $R$}}}
\put(233,585){\makebox(0,0)[lb]
    {\raisebox{0pt}[0pt][0pt]{\twlrm $h$}}}
\put(205,545){\makebox(0,0)[lb]
    {\raisebox{0pt}[0pt][0pt]{\twlrm (h)}}}
\put(315,590){\circle{40}}
\put(295,590){\circle*{8}}
\put(295,590){\line(-1, 0){ 22}}
\put(335,590){\line(1,0){ 20}}
\put(273,595){\makebox(0,0)[lb]
    {\raisebox{0pt}[0pt][0pt]{\twlrm $R$}}}
\put(345,595){\makebox(0,0)[lb]
    {\raisebox{0pt}[0pt][0pt]{\twlrm $E$}}}
\put(312,613){\makebox(0,0)[lb]
    {\raisebox{0pt}[0pt][0pt]{\twlrm $h$}}}
\put(310,545){\makebox(0,0)[lb]
    {\raisebox{0pt}[0pt][0pt]{\twlrm (i)}}}
\put(430,590){\circle{40}}
\put(410,590){\circle*{8}}
\put(410,590){\line(-1, 0){ 22}}
\put(444,604){\line(1, 1){ 15}}
\put(444,576){\line(1, -1){ 15}}
\put(388,595){\makebox(0,0)[lb]
    {\raisebox{0pt}[0pt][0pt]{\twlrm $R$}}}
\put(443,620){\makebox(0,0)[lb]
    {\raisebox{0pt}[0pt][0pt]{\twlrm $E$}}}
\put(443,550){\makebox(0,0)[lb]
    {\raisebox{0pt}[0pt][0pt]{\twlrm $E$}}}
\put(453,585){\makebox(0,0)[lb]
    {\raisebox{0pt}[0pt][0pt]{\twlrm $h$}}}
\put(405,545){\makebox(0,0)[lb]
    {\raisebox{0pt}[0pt][0pt]{\twlrm (j)}}}
\end{picture}
\caption{$O(\alpha' \kappa_0^2)$ divergent diagrams.
Internal lines with $h$ are graviton propagators.
Other internal lines are matter propagators.
Vertices with a dot are those of the one-loop counter term.}
\label{figurethree}
\vspace{5mm}
\end{figure}}
\begin{document}
\addtolength{\baselineskip}{.7mm}
\thispagestyle{empty}
\begin{flushright}
Imperial/TP/93--94/10 \\
TIT/HEP--244 \\
{\tt hep-th/9311181} \\
November, 1993
\end{flushright}
\vspace{2mm}
\begin{center}
{\Large{\bf Physical Scaling and Renormalization Group \\[2mm]
in Two-Dimensional Gravity}} \\[15mm]
{\sc Yoshiaki Tanii}\footnote{
On leave of absence from Physics Department,
Saitama University, Urawa, Saitama 338, Japan. \\
\hspace*{6.5mm}{\tt e-mail: y.tanii@ic.ac.uk}} \\[3mm]
{\it The Blackett Laboratory, Imperial College \\[2mm]
London, SW7 2BZ, U.K.}
\\[8mm]
{\sc Shin-ichi Kojima}\footnote{
\tt e-mail: kotori@phys.titech.ac.jp}
\hspace{1mm} and \hspace{1mm}
{\sc Norisuke Sakai}\footnote{
\tt e-mail: nsakai@phys.titech.ac.jp} \\[3mm]
{\it Department of Physics, Tokyo Institute of Technology \\[2mm]
Oh-okayama, Meguro, Tokyo 152, Japan} \\[15mm]
{\bf Abstract}\\[5mm]
{\parbox{13cm}{\hspace{5mm}
Quantum gravitational effects on the renormalization group
equation are studied in the $(2+\epsilon)$-dimensional approach.
Divergences in a matter one-loop effective action do not receive
gravitational radiative corrections. The renormalization factor
for beta functions recently found by Klebanov, Kogan and Polyakov
is obtained by using the renormalized cosmological constant to
define the physical scale transformation.
}}
\end{center}
\vfill
\newpage
\setcounter{section}{0}
\setcounter{equation}{0}
%
%
Recently a gravitational dressing of the renormalization group
in two spacetime dimensions
was studied in ref.\ \cite{KKP}. It was shown that
one-loop beta functions for marginal perturbations of
a conformal field theory are multiplicatively renormalized
due to quantum effects of gravity.
The beta function in the presence of quantum gravity $\tilde\beta$
is related to the beta function without quantum gravity $\beta$ as
\be
\tilde\beta = {k+2 \over k+1} \, \beta,
\label{kkpresult}
\ee
where $k$ is the level of the gravitational SL(2, R) Kac-Moody
algebra and is given in terms of the matter central charge $c$ by
\be
k = {1 \over 12} \left( c - 37 - \sqrt{(1-c)(25-c)} \right).
\ee
This result was obtained in ref.\ \cite{KKP} mainly using
the light-cone gauge for gravity \cite{KPZ}.
\par
The purpose of our paper is to study this relation further and
clarify the origin of the factor ${k+2 \over k+1}$ in
eq.\ (\ref{kkpresult}).
We will use the $(2+\epsilon)$-dimensional approach for quantum
gravity \cite{WEI}--\cite{KST} to compute quantum gravitational
effects on the beta functions. We will find that gravitational
radiative corrections to one-loop divergence in the matter effective
action cancel out at the lowest order of the gravitational constant.
Therefore one would obtain exactly the same beta functions as those
in theories without quantum gravity. We argue that the
factor in eq.\ (\ref{kkpresult}) is due to a proper definition of
the physical scale transformation in the presence of quantum gravity.
We obtain that factor by
using the dimensionless renormalized cosmological constant
to define the physical scale transformation.
We will discuss the relation (\ref{kkpresult})
in the light-cone gauge \cite{KPZ} from this point of view.
\par
%
%
We shall consider a nonlinear sigma model coupled to gravity in
$d = 2 + \epsilon$ dimensions. The action is
\be
S = \int d^d x \sqrt{-g} \left[ {1 \over 16\pi G_0} R^{(d)}
- {1 \over 4\pi\alpha'}
g^{\mu\nu} \partial_\mu X^i \partial_\nu X^j G_{0ij}(X) \right],
\label{action}
\ee
where $G_0$ is the bare gravitational constant and $G_{0ij}$ is
the bare target space metric.
Using the normal coordinate expansion \cite{NORMAL},
which we will discuss below,
we can separate this action into two parts.
The first part is the Einstein gravity coupled to free scalar
fields. The second part consists of interaction terms due to
the presence of nontrivial background fields $G_{ij}$ in the
target space.
\par
Quantization of the first part was discussed in
refs.\ \cite{KN}, \cite{KKN}. One-loop renormalization of the
gravitational constant is
\be
{1 \over G_0} = \mu^\epsilon \left( {1 \over G}
- {2 \over 3} \, {25-c \over \epsilon} \right),
\label{gravconst}
\ee
where $G$ is the renormalized gravitational constant and $\mu$ is
the renormalization scale.
It was shown in ref.\ \cite{KKN} that the two-dimensional gravity
\cite{KPZ}, \cite{DDK}
can be obtained as a limit $\epsilon \rightarrow 0$ in the
strong coupling region $|G| \gg |\epsilon|$.
In this region the bare gravitational constant is given by
\be
G_0 = - {3 \over 2} \, \mu^{-\epsilon} {\epsilon \over 25-c}.
\label{gzero}
\ee
Using the Einstein action with this bare gravitational constant,
anomalous dimensions of physical operators were obtained and were
shown to be consistent with the previous results using other
methods \cite{KPZ}, \cite{DDK}.
\par
Using this approach for quantum gravity
we can now study effects of quantum gravity on the beta functions
of matter self-couplings by taking the second part of the action
into account. There are two parameters for loop expansions.
The parameter for matter loops is $\alpha'$ and that
for graviton loops is $G_0$ in eq.\ (\ref{gzero}).
Effects of quantum gravity become small for large $-c$.
In the limit $c \rightarrow -\infty$ we obtain a theory without
quantum gravity.
\par
To obtain the beta functions at the first order in $\alpha'$
we compute divergences in the effective action using the matter
action in eq.\ (\ref{action}) with $G_{0ij}$ replaced by
$\mu^\epsilon G_{ij}$, where $G_{ij}$ is the dimensionless
renormalized coupling. We use the background field method
\cite{DEWITT} and the normal coordinate expansion \cite{NORMAL}.
The matter fields are expanded as $X^i = \X^i + \xi^i + O(\xi^2)$,
where $\X^i$ are background fields and $\xi^i$ are quantum
fluctuations (normal coordinates). It is more convenient to use
$\xi^a = \xi^i E_i{}^a(\X)$ as quantum fields, where $E_i{}^a$
is the target space vielbein: $G_{ij} = E_i{}^a E_j{}^b \eta_{ab}$.
The matter action can be expanded as \cite{NORMAL}
\ba
S_{\rm matter}
\A = \A - \, {\mu^\epsilon \over 4\pi\alpha'} \int d^d x \sqrt{-g}
g^{\mu\nu} \biggl[ \partial_\mu \X^i \partial_\nu \X^j
G_{ij}(\X)
+ 2 D_\mu \xi^a \partial_\nu \X^i E_{ia}(\X)  \nonu
\A \A + \, D_\mu \xi^a D_\nu \xi^b \eta_{ab}
- \partial_\mu \X^i \partial_\nu \X^j R_{iajb}(\X)
\xi^a \xi^b \nonu
\A \A - \, {4 \over 3} \partial_\mu \X^i \xi^a \xi^b D_\nu \xi^c
R_{caib}(\X)
- {1 \over 3} \partial_\mu \X^i \partial_\nu \X^j
\xi^a \xi^b \xi^c D_c R_{iajb}(\X) \nonu
\A \A - \, {1 \over 3} D_\mu \xi^a D_\nu \xi^b \xi^c \xi^d
R_{acbd}(\X)
- {1 \over 2} \xi^a \xi^b \xi^c D_\mu \xi^d \partial_\nu \X^i
D_c R_{daib}(\X) \nonu
\A \A - \, {1 \over 12} \partial_\mu \X^i \partial_\nu \X^j
\xi^a \xi^b \xi^c \xi^d \left( D_a D_b R_{icjd}
- 4 R^k{}_{aib} R_{kcjd} \right)(\X) + O(\xi^5) \biggr].
\label{normalexp}
\ea
The covariant derivative is defined as
\be
D_\mu \xi^a = \partial_\mu \xi^a
+ \partial_\mu \X^i \Omega_i{}^a{}_b(\X) \xi^b,
\label{covder}
\ee
where $\Omega_i{}^a{}_b$ is the target space spin connection.
\par
The metric is parametrized by a background field $\hat g_{\mu\nu}$
and quantum fields $h^{\mu}{}_\nu$, $\phi$ as \cite{KKN}
\be
g_{\mu\nu} = \hat g_{\mu\rho} \, (\e^{\kappa_0 h})^\rho{}_\nu
\e^{- \kappa_0 \phi}, \quad
h_{\mu\nu} \equiv \hat g_{\mu\rho} h^\rho{}_\nu = h_{\nu\mu}, \quad
h^\mu{}_\mu = 0,
\label{metric}
\ee
where $\kappa_0^2 = 16\pi G_0$.
The metric dependence of the matter action is
\be
\sqrt{-g} g^{\mu\nu}
= \sqrt{-\hat g} \, (\e^{- \kappa_0 h})^\mu{}_\rho \,
\hat g^{\rho\nu} \e^{-{1 \over 2}\epsilon\kappa_0 \phi}.
\label{metricdep}
\ee
Notice that the conformal mode $\phi$ is multiplied by a factor
$\epsilon$.
This corresponds to the fact that the perturbation due to
the nontrivial background $G_{ij}$ is a marginal operator
in two dimensions.
We use the gauge fixing term of ref.\ \cite{KKN}
for the general coordinate symmetry
\be
S_{\rm GF} = - {1 \over 2} \int d^d x \sqrt{-\hat g}
\left( \hat D^\nu h_{\nu\mu}
+ {1 \over 2} \epsilon \hat D_\mu \phi \right)^2.
\ee
In this gauge the propagators of the gravitational fields are
\ba
\VEV{h_{\mu\nu}(x) h_{\rho\sigma}(y)}
\A = \A - i \left( \eta_{\mu\rho} \eta_{\nu\sigma}
+ \eta_{\mu\sigma} \eta_{\nu\rho}
- {2 \over d} \eta_{\mu\nu} \eta_{\rho\sigma} \right)
\int {d^d p \over (2\pi)^d} {1 \over p^2}
\e^{i p \cdot (x-y)}, \nonu
\VEV{\phi(x) \phi(y)}
\A = \A {4i \over \epsilon d} \int {d^d p \over (2\pi)^d}
{1 \over p^2} \e^{i p \cdot (x-y)}, \qquad
\VEV{h_{\mu\nu}(x)\phi(y)} = 0.
\label{prop}
\ea
\par
First let us recall the one-loop beta function when gravity is
not quantized \cite{FRIEDAN}, \cite{NORMAL}.
We are interested in order
$O(\alpha')$ terms in the effective action $\Gamma$.
As shown in Fig.\ 1, there are three divergent diagrams of this order.
However, the divergence of the diagram (b)
is canceled by that of the diagram (c). They should cancel since
they are not allowed by the covariance and the local Lorentz symmetry
in the target space.
Only the diagram (a) contributes to
one-loop divergence. We obtain
\figone
\be
\Gamma = - {1 \over 4\pi\alpha'} {\alpha' \zeta \over \epsilon}
\int d^d x \sqrt{-\hat g} \hat g^{\mu\nu}
\partial_\mu \X^i \partial_\nu \X^j R_{ij}(\X)
+ O(\epsilon^0),
\label{divergence}
\ee
where $\zeta = 1$.
The divergence can be removed by adding a counter term
\be
S_{\rm c.t.} = {\mu^\epsilon \over 4\pi\alpha'}
{\alpha' \over \epsilon} \int d^d x \sqrt{-g} g^{\mu\nu}
\partial_\mu X^i \partial_\nu X^j R_{ij}(X)
\label{counterterm}
\ee
to the matter action. This counter term corresponds to
choosing the bare coupling in eq.\ (\ref{action}) as
\be
G_{0ij} = \mu^\epsilon \left( G_{ij}
- {\alpha' \over \epsilon} R_{ij} \right).
\label{bare}
\ee
{}From the $\mu$-independence of the bare coupling $G_{0ij}$ we obtain
the one-loop beta function in the limit $\epsilon \rightarrow 0$ as
\be
\beta_{ij} \equiv \mu {\partial \over \partial\mu} G_{ij}
= \alpha' R_{ij}.
\label{oneloopbeta}
\ee
\par
Now we shall consider quantum effects of gravity.
At order $O(\kappa_0^2)$ there are two divergent diagrams
as shown in Fig.\ 2.
In these diagrams the internal lines are $h_{\mu\nu}$ or $\xi^a$.
Diagrams with internal $\phi$ lines do not contribute to divergences
because the $\phi$-vertices always contain a factor $\epsilon$
as shown in eq.\ (\ref{metricdep}).
Their divergences have the form
\figtwo
\be
\Gamma =  - {1 \over 4\pi\alpha'} {\eta \over \epsilon}
\int d^d x \sqrt{- \hat g}
\hat g^{\mu\nu} \partial_\mu \X^i \partial_\nu \X^j G_{ij}(\X).
\label{kappadiv}
\ee
The coefficient $\eta$ for each diagram is given by
\be
\eta_{(a)} = - {\kappa_0^2 \over 2\pi}, \qquad
\eta_{(b)} = {\kappa_0^2 \over 2\pi}.
\label{etavalue}
\ee
Therefore the divergences of these two diagrams cancel each other
and there is no divergence of order $O(\kappa_0^2)$.
Actually, the coefficients in eq.\ (\ref{etavalue}) are of order
$O(\epsilon)$ since $\kappa_0^2 = O(\epsilon)$, and
eq.\ (\ref{kappadiv}) for each diagram is finite.
\par
Next let us consider order $O(\alpha' \kappa_0^2)$ divergences.
We have to compute divergences in two-loop
diagrams. Since $\kappa_0^2 = O(\epsilon)$, we only need to know
$\epsilon^{-2}$ singularities of Feynman integrals.
We have to take into account the vertices in the one-loop
counter term (\ref{counterterm}).
The normal coordinate expansion of this counter term is
\ba
S_{\rm c.t.}
\A = \A {\mu^\epsilon \over 4\pi\alpha'}
{\alpha' \over \epsilon} \int d^d x \sqrt{-g} g^{\mu\nu}
\biggl[ \partial_\mu \X^i \partial_\nu \X^j R_{ij}(\X)
+ \partial_\mu \X^i \partial_\nu \X^j \xi^a D_a R_{ij}(\X) \nonu
\A \A + 2 \partial_\mu \X^i D_\nu \xi^a R_{ia}(\X)
+ {1 \over 2} \partial_\mu \X^i \partial_\nu \X^j \xi^a \xi^b
\left( D_a D_b R_{ij} - 2 R^k{}_{a i b} R_{kj} \right)
(\X) \nonu
\A \A + 2 \partial_\mu \X^i D_\nu \xi^a \xi^b D_b R_{ia}(\X)
+ D_\mu \xi^a D_\nu \xi^b R_{ab}(\X) + O(\xi^3) \biggr].
\ea
Divergent diagrams of order $O(\alpha' \kappa_0^2)$ are shown
in Fig.\ 3.
In addition to the diagrams in Fig.\ 3 there are divergent
diagrams with the $\Omega$ vertices.
However, they should cancel each other by the covariance and
the local Lorentz symmetry in the target space as in Fig.\ 1.
Divergence in each diagram in Fig.\ 3 has the same form as
eq.\ (\ref{divergence}). The coefficient $\zeta$ for each
diagram is
\figthree
\ba
\zeta_{\rm (a)} \A = \A 0, \quad
\zeta_{\rm (b)} = - {\kappa_0^2 \over 2\pi\epsilon}, \quad
\zeta_{\rm (c)} = {\kappa_0^2 \over 2\pi\epsilon}, \quad
\zeta_{\rm (d)} = - {\kappa_0^2 \over 2\pi\epsilon}, \quad
\zeta_{\rm (e)} = {2 \kappa_0^2 \over 3 \pi \epsilon}, \nonu
\zeta_{\rm (f)} \A = \A - {\kappa_0^2 \over 6\pi\epsilon}, \quad
\zeta_{\rm (g)} = 0, \quad
\zeta_{\rm (h)} = {\kappa_0^2 \over 2\pi\epsilon}, \quad
\zeta_{\rm (i)} = - {\kappa_0^2 \over \pi\epsilon}, \quad
\zeta_{\rm (j)} = {\kappa_0^2 \over 2\pi\epsilon}.
\label{zetavalue}
\ea
Since $\kappa_0^2 = O(\epsilon)$, these coefficients are finite
for $\epsilon \rightarrow 0$. We see that the sum of
divergences in the diagrams (a)--(j) cancel out.
This cancellation is similar to the observation in ref.\ \cite{KKN}
that the diagrams with $h_{\mu\nu}$ lines do not contribute to the
renormalization of the operator $\e^{-{d \over 2}(1-\Delta_0)\phi}$
to two-loop order. In the present case of marginal operators
there is no $\phi$ contribution either. It can happen that
gravitational corrections to divergences completely cancel
also at higher orders of the loop expansions in $\kappa_0^2$.
This may not be so surprising because $h_{\mu\nu}$ is a gauge degree
of freedom in two dimensions and can be set to zero by a general
coordinate transformation.
\par
Thus we have seen that the one-loop divergence does not
receive gravitational radiative correction.
Therefore we obtain the same beta function (\ref{oneloopbeta})
as in the theory without quantum gravity.
This is consistent with the result in ref.\ \cite{KKP} in the
sense that the beta functions are essentially the same in theories
with and without quantum gravity. The factor ${k+2 \over k+1}$ in
eq.\ (\ref{kkpresult}) can be understood
by considering a definition of the physical scale transformation.
\par
%
%
Without gravitational interactions, the beta function
(\ref{oneloopbeta}) is defined as a response of the renormalized
quantities to the change of the renormalization scale $\mu$.
On the other hand, the characteristic feature of gravity is to provide
a metric which can determine the physical scale in the theory.
Therefore we should measure the response of the renormalized quantities
to the change of the physical scale.
It has been pointed out that the renormalization of the gravitational
constant $G$ can be made meaningful only after fixing the scale of the
metric by renormalizing a reference operator $O$ \cite{KN}.
Since the gravitational constant becomes dimensionless in
two-dimensions, it is most natural to consider the reference operator
which also has dimensionless coupling in two dimensions.
Namely we choose the reference operator to be spinless and with a
conformal dimension $(1, 1)$, such as the Thirring interaction.
After fixing the scale of the metric, we can absorb divergences
of the type  $\sqrt{-g} R$ into a renormalization of
the gravitational constant $G$, and the divergences of the type
$\sqrt{-g}$ into a renormalization of the cosmological constant
$\Lambda$.
Instead of the gravitational constant, the cosmological constant
$\Lambda$ naturally provides the physical scale in the presence of
the gravitational interaction in two dimensions, since it is the only
dimensionful parameter in the gravity theory in two dimensions.
\par
To use the cosmological term in defining the physical scale
transformation, let us introduce the cosmological term
into the action (\ref{action})
\be
\Lambda_0 \int d^d x \sqrt{-g},
\ee
where $\Lambda_0$ is the bare cosmological constant.
Renormalization of the cosmological term was discussed in
ref.\ \cite{KKN}.
To remove divergences, the bare cosmological
constant is expressed as
\be
\Lambda_0 = \mu^d \Lambda Z,
\label{rencosm}
\ee
where $Z$ is a divergent renormalization factor and
$\Lambda$ is a dimensionless renormalized cosmological constant.
{}From the $\mu$-independence of $\Lambda_0$ we obtain
\be
\mu {\partial \Lambda \over \partial \mu}
= - ( d + \gamma ) \Lambda, \qquad
\gamma = \mu {\partial \over \partial \mu} \ln Z,
\label{cosmreneq}
\ee
where $\gamma$ is the anomalous dimension. In the two-dimensional
limit the exact form of $\gamma$ was obtained in ref.\ \cite{KKN}
by considering divergences coming from the conformal mode
to arbitrary orders in the loop expansion
\be
\gamma = - 2 - \alpha Q,
\ee
where $\alpha$ and $Q$ are defined by
\be
Q = \sqrt{25 - c \over 3}, \qquad
\alpha = - {1 \over 2\sqrt{3}} \left( \sqrt{25-c}
- \sqrt{1-c} \right).
\label{ddkparameter}
\ee
In this case the solution of eq.\ (\ref{cosmreneq}) is
\be
\Lambda \sim \mu^{-{2 \over x}}, \qquad
{1 \over x} = 1 + {\gamma \over 2} = - {\alpha Q \over 2}.
\label{cosmmudep}
\ee
In the limit of no quantum gravity $c \rightarrow -\infty$
the anomalous dimension vanishes $\gamma \rightarrow 0$.
\par
Now let us consider the beta function in the presence of gravity.
Instead of the renormalization scale $\mu$,
we use the physical scale derived from the cosmological constant
to define the beta function.
Since the renormalized cosmological constant $\Lambda$ is defined
by eq.\ (\ref{rencosm}), the inverse
square root of the cosmological constant $\sqrt{\Lambda^{-1}}$
should be used in place of $\mu$ in $d = 2$ dimensions.
Without gravitational interactions, the beta function is usually
defined as a change of the renormalized
coupling $G_{ij}$ in response to the change of the renormalization scale
\be
\beta_{ij}
= {\partial G_{ij}(\mu) \over \partial\ln\mu}.
\label{betafunction}
\ee
In the presence of gravity, we should consider the beta function
as a response to the physical scale $\sqrt{\Lambda^{-1}}$
\be
\tilde \beta_{ij}
= {\partial G_{ij} \over \partial \ln \sqrt{\Lambda(\mu)^{-1}}}
= {\partial \ln \mu \over \partial \ln \sqrt{\Lambda(\mu)^{-1}}}
{\partial G_{ij} \over \partial \ln \mu}
= x \beta_{ij}.
\label{gravbetafunction}
\ee
In the limit of no quantum gravity $c \rightarrow -\infty$, we have
$x \rightarrow 1$ and $\tilde\beta_{ij} \rightarrow \beta_{ij}$.
Using eqs.\ (\ref{cosmmudep}) and (\ref{ddkparameter}) we see
that the relation between $\tilde\beta_{ij}$ and $\beta_{ij}$
in eq.\ (\ref{gravbetafunction}) is exactly the same as
eq.\ (\ref{kkpresult}).
\par
Another way to state this scale transformation is the following.
We can consider a generalized renormalization scale transformation
$\mu \rightarrow \lambda^y \mu $ with a power $y$ of the
scaling parameter $\lambda$. We would like to require
the renormalized cosmological constant $\Lambda$ in eq.\ (\ref{rencosm})
to have the classical response in two dimensions
\be
\Lambda \rightarrow \lambda^{-2} \Lambda
\label{cosmchange}
\ee
even in the presence of quantum effects.
Therefore the renormalization scale dependence in eq.\ (\ref{cosmmudep})
dictates $y=x$.
Clearly this transformation assures that the scale transformation
of the cosmological constant be independent of the matter content of
the theory.
We can now define the beta function in the presence of quantum gravity
as a response to this physical scale transformation
\be
\tilde\beta_{ij} = \lim_{\lambda \rightarrow 1}
{1 \over \ln\lambda} \bigl[
G_{ij}(\lambda^x \mu) - G_{ij}(\mu) \bigr]
= x \beta_{ij},
\ee
where $\beta_{ij}$ is defined in eq.\ (\ref{betafunction}).
Thus we obtain the same $\tilde\beta_{ij}$ as in
eq.\ (\ref{gravbetafunction}).
\par
Let us comment on modifications if one chooses to use
other physical operators instead of the
cosmological term to define a physical scale transformation of $\mu$.
We can repeat the above analysis using the result on the
renormalization of physical operators in ref.\ \cite{KKN}.
If we use a physical operator corresponding to a matter operator
with a conformal dimension $(\Delta_0, \Delta_0)$, we obtain a relation
\be
\tilde\beta_{ij} = - {2(1-\Delta_0) \over \beta Q} \beta_{ij},
\ee
where
\be
\beta = - {1 \over 2\sqrt{3}} \left( \sqrt{25-c}
- \sqrt{1-c+24\Delta_0} \right).
\ee
However, we would like to emphasize that the usual scaling arguments
in the presence of gravity are in fact based upon the use of
the physical scale defined in terms of the cosmological constant
$\Lambda$ \cite{DDK}.
We are here advocating a similar argument to define the
beta function in the presence of gravity.
\par
%
%
To confirm the above interpretation of the factor in
eq.\ (\ref{kkpresult}) we shall examine the physical scaling
in the light-cone gauge \cite{KPZ}. As in ref.\ \cite{KKP}
the beta functions can be obtained from coefficients of the
operator product expansion (OPE), which in turn can be obtained
from a ratio of two- and three-point correlation functions.
To see how the cutoff and the renormalization scale are
introduced in this method let us recall the
relation between the OPE coefficients and the beta
functions. As in ref.\ \cite{KKP} we consider an action
for a conformal field theory perturbed by
marginal (conformal weight one) operators $O_n(x)$
\be
S = S_0 + \sum_n \lambda_n O_n, \qquad
O_n = \int d^2 x \, O_n(x).
\ee
The vacuum expectation value can be expanded as
\be
\VEV{\cdots} =
\VEV{(\cdots) \left( 1 +  i \sum_n \lambda_n O_n
+ {1 \over 2} i^2 \sum_{n,m} \lambda_n \lambda_m O_n O_m
+ \cdots \right)}_0.
\ee
Let us consider the $O(\lambda^2)$ terms.
The $x$-integrations in $O_1$ and $O_2$ give short distance
divergences.
The short distance behavior
of the product of the integrands is given by the OPE
\be
O_n(x_1) O_m(x_2)
\sim {1 \over (x_1-x_2)^2} \sum_l g_l{}^{nm} \, O_l(x_2).
\ee
Therefore the $x$-integrations give
\ba
O_n O_m
\A = \A \int d^2 x_1 d^2 x_2 {1 \over (x_1 - x_2)^2 + \epsilon^2}
\sum_l g_l{}^{nm} O_l(x_2) \nonu
\A = \A 2 \pi i \ln\epsilon \sum_l g_l{}^{nm} O_l + \cdots,
\label{oodiv}
\ea
where we have regularized the integral by making a Wick rotation
$x^0 = -i x^2$ and introducing a cutoff $\epsilon$
in the coordinate space.
This short distance divergence can be removed by introducing
the bare coupling constants
\be
\lambda_{0n} = \lambda_n + \pi \ln (\epsilon\mu)
\sum_{l,m} \lambda_l \lambda_m g_n{}^{lm} + O(\lambda^3),
\ee
where we have introduced a renormalization scale $\mu$.
Then the beta functions are given by
\be
\beta_n
= \mu {\partial \lambda_n \over \partial \mu}
= - \pi \sum_{l,m} \lambda_l \lambda_m g_n{}^{lm} + O(\lambda^3).
\label{opebeta}
\ee
Thus we can obtain the beta functions from
the OPE coefficients $g_n{}^{lm}$.
\par
%
%
In the light-cone gauge the metric has the form
\be
g_{\mu\nu} dx^\mu dx^\nu = - 2 dx^+ dx^- + h_{++} dx^+ dx^+
\ee
and the cosmological term is a c-number
\be
\Lambda_0 \int d^2 x \sqrt{-g} = \Lambda_0 \int d^2 x.
\ee
There is no short distance divergence.
Therefore the relation between the bare
and dimensionless renormalized cosmological constants is
$\Lambda_0 = \mu^2 \Lambda$ and we obtain
\be
\Lambda \sim \mu^{-2}.
\ee
We see that the usual scaling $\mu \rightarrow \lambda \mu$
is physical, i.e., it changes the renormalized
cosmological constant $\Lambda$ as in eq.\ (\ref{cosmchange}).
We should obtain the factor in eq.\ (\ref{kkpresult}) for the
usual beta functions.
Indeed it was obtained in ref.\ \cite{KKP} in this way.
Therefore our interpretation of the factor is consistent with
the analysis in the light-cone gauge.
\par
The gravitational dressing of the beta function in the conformal
gauge has already been discussed in ref.\ \cite{KKP} using the
cosmological constant
as the physical scale similarly to our treatment.
The renormalization group equation in the conformal gauge was
also discussed in ref.\ \cite{SCH}.
\par
%
%
\vspace{5mm}
One of the authors (Y.T.) would like to thank C.M. Hull and
K.S. Stelle for useful discussions.
He also would like to thank the Theoretical
Physics Group of Imperial College for hospitality, and the
Japan Society for the Promotion of Science and the Royal Society
for a grant.
This work is supported in part by Grant-in-Aid for Scientific
Research (S.K.) and (No.05640334) (N.S.), and Grant-in-Aid for
Scientific Research for Priority Areas (No.05230019) (N.S.) from
the Ministry of Education, Science and Culture.
%
%
\newpage
\newcommand{\NP}[1]{{\it Nucl.\ Phys.\ }{\bf #1}}
\newcommand{\PL}[1]{{\it Phys.\ Lett.\ }{\bf #1}}
\newcommand{\CMP}[1]{{\it Commun.\ Math.\ Phys.\ }{\bf #1}}
\newcommand{\MPL}[1]{{\it Mod.\ Phys.\ Lett.\ }{\bf #1}}
\newcommand{\IJMP}[1]{{\it Int.\ J.\ Mod.\ Phys.\ }{\bf #1}}
\newcommand{\PR}[1]{{\it Phys.\ Rev.\ }{\bf #1}}
\newcommand{\PRL}[1]{{\it Phys.\ Rev.\ Lett.\ }{\bf #1}}
\newcommand{\PTP}[1]{{\it Prog.\ Theor.\ Phys.\ }{\bf #1}}
\newcommand{\PTPS}[1]{{\it Prog.\ Theor.\ Phys.\ Suppl.\ }{\bf #1}}
\newcommand{\AP}[1]{{\it Ann.\ Phys.\ }{\bf #1}}
\end{document}